\documentclass[pdflatex,sn-basic]{sn-jnl}

\usepackage{natbib}
\usepackage{xspace}
\usepackage{tabularx}
\usepackage{hyperref}
\usepackage{graphicx}

\jyear{2022}%

\def\Hers{\emph{Herschel}}

\def\OAS{\emph{OASIS}}
\def\jwst{\emph{JWST}}
\def\hyd{{H$_{2}$}}
\def\wat{{H$_{2}$O}}

\def\watisoA{{H$_{2}^{16}$O}}
\def\watisoB{{H$_{2}^{18}$O}}

\def\hh{{H$_{2}$}}
\def\um{{$\mu$m}}

\def\kms{{km s$^{-1}$}}
\def\mj{{M$_\mathrm{J}$}}
\newcommand{\ee}[1]{\mbox{${} \times 10^{#1}$}}

\usepackage{titlesec}

\titleformat{\section}
  {\normalfont\large\bfseries}{\thesection}{0.5em}{}
\titleformat{\subsection}
  {\normalfont\normalsize\bfseries}{\thesubsection}{0.5em}{}
\titleformat{\subsubsection}
  {\normalfont\normalsize\bfseries}{\thesubsubsection}{0.5em}{}
  \titleformat{\paragraph}
  {\normalfont\normalsize\bfseries}{\theparagraph}{0.5em}{}
\titleformat*{\subparagraph}{\normalsize\bfseries\sffamily}

\begin{document}

\title[Protoplanetary Disk Science with OASIS]{Protoplanetary Disk Science with the Orbiting Astronomical Satellite Investigating Stellar Systems (OASIS) Observatory}

\author*[1]{\fnm{Kamber R.} \sur{Schwarz}}\email{schwarz@mpia.de}
\author[2]{\fnm{Joan} \sur{Najita}}\email{joan.najita@noirlab.edu}
\author[3]{\fnm{Jennifer} \sur{Bergner}}\email{jbergner@uchicago.edu}
\author[4]{\fnm{John} \sur{Carr}}\email{jscarr108@gmail.com }
\author[5]{\fnm{Alexander} \sur{Tielens}}\email{tielens@strw.leidenuniv.nl}
\author[6]{\fnm{Edwin A.} \sur{Bergin}}\email{ebergin@umich.edu}
\author[7]{\fnm{David} \sur{Wilner}}\email{dwilner@cfa.harvard.edu }
\author[8]{\fnm{David} \sur{Leisawitz}}\email{david.t.leisawitz@nasa.gov}
\author[9]{\fnm{Christopher K.} \sur{Walker}}\email{iras16293@gmail.com}

\affil*[1]{\orgname{Max Planck Institute for Astronomy}, \orgaddress{\street{K\"{o}nigstuhl 17}, \city{Heidelberg}, \country{Germany}}}

\affil[2]{\orgname{NSF's NOIRLab}, \orgaddress{\street{950 N. Cherry Avenue}, \postcode{85719}, \state{AZ}, \country{USA}}}

\affil[3]{\orgname{Department of Geophysical Sciences}, \orgaddress{\street{University of Chicago}, \postcode{60637}, \state{IL}, \country{USA}}}

\affil[4]{\orgname{Astronomy Department}, \orgaddress{\street{University of Maryland}, \postcode{20742}, \state{MD}, \country{USA}}}

\affil[5]{\orgname{Leiden Observatory}, \orgaddress{\street{P.O. Box 9513}, \city{Leiden}, \country{The Netherlands}}}

\affil[6]{\orgname{Department of Astronomy}, \orgaddress{\street{University of Michigan}, \postcode{48109}, \state{MI}, \country{USA}}}

\affil[7]{\orgname{Center for Astrophysics Harvard \& Smithsonian}, \orgaddress{\street{60 Garden Street}, \postcode{02138}, \state{MA}, \country{USA}}}

\affil[8]{\orgname{NASA Goddard Space Flight Center}, \orgaddress{\street{800 Greenbelt Road}, \postcode{20771}, \state{MD}, \country{USA}}}

\affil[9]{\orgname{Department of Astronomy and Steward Observatory}, \orgaddress{\street{University of Arizona}, \postcode{85719}, \state{AZ}, \country{USA}}}

\abstract{The \emph{Orbiting Astronomical Satellite for Investigating Stellar Systems (OASIS)} is a NASA Astrophysics MIDEX-class mission concept, with the stated goal of \emph{Following water from galaxies, through protostellar systems, to Earth's oceans}. This paper details the protoplanetary disk science achievable with \OAS. \OAS's suite of heterodyne receivers allow for simultaneous, high spectral resolution observations of water emission lines spanning a large range of physical conditions within protoplanetary disks. These observations will allow us to map the spatial distribution of water vapor in disks across evolutionary stages and assess the importance of water, particularly the location of the midplane water snowline, to planet formation. \OAS\ will also detect the H$_2$ isotopologue HD in 100+ disks, allowing for the most accurate determination of total protoplanetary disk gas mass to date. When combined with the contemporaneous water observations, the HD detection will also allow us to trace the evolution of water vapor across evolutionary stages. These observations will enable \OAS\ to characterize the time development of the water distribution and the role water plays in the process of planetary system formation.} 


\maketitle

\section{Strategic Motivation}
\label{sec:motivation}  
Water is key to the emergence of life on Earth. Understanding the origin of Earth's water is thus a major goal of modern astronomy. Observations of water from the ground, however, are extremely challenging due to the strong water absorption lines in the Earth's atmosphere. Studying extra-terrestrial water therefore necessitates the use of space-borne observatories. Low energy transitions of water, such as those originating from the cold ($\sim 10-1000$ K) planet forming disks around young stars, emit at submillimeter to far-infrared (IR) wavelengths. Thus, any mission aiming to understand 
how and from where (exo)planets like Earth obtain their water
must cover these wavelengths. There is currently no space-borne facility capable of observing low energy water emission. The \emph{Orbiting Astronomical Satellite Investigating Stellar Systems (OASIS)} observatory is an Astrophysics MIDEX-class mission concept specifically designed to \emph{follow the trail of water from galaxies to protoplanetary disks to our solar system}.

\OAS\ combines a 14 meter inflatable reflector with a suite of
heterodyne receivers covering the frequency ranges 0.45-0.58, 1.1-2.2, 2.5-2.9, and 4.7 THz. 
This observatory will enable high spectral resolution (R $\ge 10^{5} $ or $\le$ 1 \kms) observations at greater than ten times the sensitivity of any previous far-infrared mission capable of observing water. For a full description of the technical capabilities of \OAS\ see \citet{Walk2021}. 

\section{The role of water in protoplanetary disk evolution and planet formation}
\label{sec:DH}
 The question of how the Earth, and by extension other terrestrial (exo-)planets, obtained its water is heavily debated. Water may have been acquired either through the accretion of volatile-rich planetesimals from the colder regions of the Earth’s embryonic feeding zone during its build up \citep{Greenberg78,Kokubo98}, or it may have been delivered to the Earth by the capture of pebbles drifting through the inner solar system on their journey to the Sun \citep{Ida19}, or the addition of water may have been a late veneer acquired via impacts of asteroids or comets after planet formation was mostly complete \citep{Morbidelli00,Owen92}.  

Planet formation begins with 
the gravitational collapse of a molecular cloud core to 
form
a central protostellar object, which, because of angular momentum conservation, is surrounded by an accretion disk. This system is embedded in an envelope of infalling material \citep{Shu87}. Several stages can be discerned in the subsequent evolution \citep{Andre00}. During the earliest phase – the Class 0 phase – the protostellar object is deeply embedded in the envelope and still actively accreting mass at a high rate from the envelope through the disk. Some of the accretion energy is converted into a stellar jet and/or disk wind dispersing the reservoir of material in the envelope.  As the envelope mass decreases due to ongoing accretion and dispersal by stellar wind, the object will enter the Class I phase, typically after some 100,000–200,000 years. During this late accretion phase, the envelope mass will continue to decrease and eventually the system will enter the Class II phase of an optically visible classical T Tauri star surrounded by a rotating protoplanetary disk. Planet formation is thought to occur during the Class I and Class II phases \citep[e.g.,][]{Drazkowska22}. Finally, as accretion onto the central star drops and the gas disk dissipates, the system transitions to the Class III phase of a weak-line T Tauri star, harboring a nascent planetary system.

\begin{figure}[ht]
   \begin{center}
   \includegraphics[width=\textwidth]{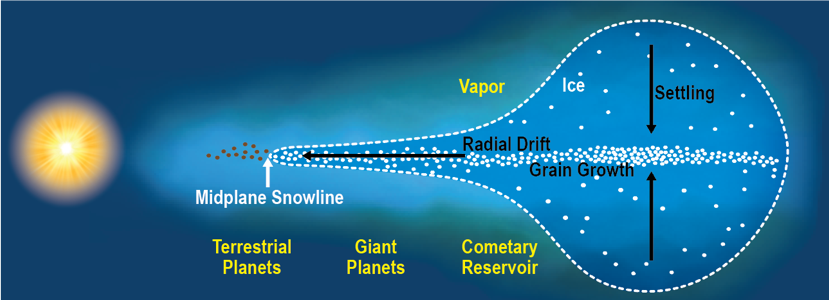}
   \end{center}
   \vspace{0mm}
  \caption{Cartoon cross section of a protoplanetary disk. In the inner disk and the upper layers of the outer disk water is in the gas phase. In the cool, outer regions of the disk, water is frozen out as ice on the surface of dust grains. The zone where water transitions between existing primarily in the gas versus the ice is called the water snowline. The snowline at the midplane, where planet formation occurs, is thought to separate the gas giant and terrestrial planet forming regions. \label{fig:cartoon}}
\end{figure}

Water plays a central role in planet formation. When fully condensed in icy form, water enhances the solid mass available for planet formation, and, therefore, the planet formation efficiency \citep{Hayashi81,Poll1996}. 
In the initial planet formation phase, icy dust grains are expected to settle out of the atmospheres of protoplanetary disks and collide, growing into larger solids. The layer of water ice increases the sticking efficiency of colliding grains, thus enhancing grain growth. As these icy grains settle, they remove water from the disk surface, dehydrating the atmosphere \citep[][Figure~\ref{fig:cartoon}]{Bergin10,Hogerheijde11,Krijt16b,Krijt20}.
The growing icy pebbles migrate inward, enhancing the solid surface density at smaller radii, which, assisted by instabilities, triggers the formation of planetesimals, the
building blocks of planets via, e.g., the streaming instability. 

The growth and accumulation of icy planetesimals at radii larger than the snowline---the locus where water transitions from 
gaseous to icy
form---fuels the formation of planets: ice giants, gas giants, and even terrestrial-sized icy planets. These water rich planets can migrate inward, interior to the midplane water snowline, to the habitable zone, bringing along with them their own supply of water. The compact millimeter continuum emission of most protoplanetary disks \citep{Long19}, along with the recognition that many exoplanetary systems are much more compact than our own Solar System and incorporate large solid masses at small orbital distances \citep{Chiang13}, may reflect the outcome of the inward migration process of both dust grains and planets. In contrast to this formation pathway, other planets may form ``dry'' within the midplane snowline, without built-in water reservoirs.  Icy planetesimals at greater distances from the host star (i.e., comets and asteroids) that are “leftover” from the planet formation process may later bombard these planets and deliver water and prebiotic molecules to their surfaces, fostering their habitability.

\section{\OAS\ Relevance to Protoplanetary Disk Science}
\label{sec:relevance}

The \OAS\ protoplanetary disk science objective is to \emph{characterize the time 
evolution
of the water distribution and the role water plays in the process of planetary system formation.}
Understanding the distribution of water in protoplanetary disk systems, the processes that regulate its abundance and its evolution, as well as its relationship to the properties of the dust in these disks and the forming planets therein is a key question in astrophysics. \OAS\ will measure the water content of a broad sample of protoplanetary disks whose structure and properties will have been characterized by ALMA and \jwst. The high spectral resolution will allow \OAS\ to investigate the spatial distribution of water through Doppler tomography using a set of carefully selected lines that probe the wide range of physical conditions inherent to protoplanetary disks (Section~\ref{sec:distribution}). The observed sample of disks will include systems spanning a wide range of ages to measure the time evolution of the water content of protoplanetary disks and trace its role in ongoing planet formation. In addition to water, \OAS\ covers multiple spectral features from more complex carbon, hydrogen, oxygen, nitrogen, phosphorous, and sulfur bearing molecules. The prebiotic science capabilities of \OAS\ are discussed in detail by \citet{Bergner22}.

The recent Astrophysics Decadal Survey \citep[\emph{Pathways to Discoveries in Astronomy and Astrophysics for the 2020s;}][]{astro2020} highlights the need to study the epoch of planet formation. \OAS\ addresses Decadal Question F-Q4: \emph{Is planet formation fast or slow?} by investigating nascent planet formation. Specifically, \OAS\ will trace the disappearance of water from the gas phase as water condenses out and fuels planet formation by measuring the distribution of total gas mass and water abundance in disk atmospheres over a range of disk radii. Previous work with \Hers\ has revealed tantalizing, but tentative and limited evidence of water condensation. The \Hers\ WISH program found that gaseous water emission from protoplanetary disks was under-abundant compared to expectations \citep{Du17,vanDishoeck21}. The large number of water emission upper limits from the WISH program may imply that the formation of large solids and the accompanying dehydration and migration process is well advanced in Class II protoplanetary disks. \OAS\ will confirm and quantify the extent of this process. \emph{OASIS}’ $>16\times$ improvement in sensitivity relative to \Hers\ will make observations of water in disks routine and enable surveys with a statistically
significant number of samples. 

\section{Overview of \OAS\ Protoplanetary Disk Science}

\subsection{Probing Planet Formation with Water}
\OAS\ will examine the first steps of planet formation using measurements of water vapor and total gas mass.
The observable water vapor in the cold outer disk is due to
water ice that is photo-desorbed from solid surfaces and 
lofted to larger heights via vertical diffusion \citep[e.g.,][]{Krijt16b}.
One possible interpretation of the low WISH water line fluxes is that 
the water vapor distribution is more compact than the gaseous disk as
traced by CO, because the inward migration of icy pebbles has
dehydrated the outermost radii. 
If true, the dominant emission radius of gaseous water will
correlate with the size of the millimeter dust disk as traced by continuum
emission, reflecting the inward radial migration of icy solids that 
fuels planetesimal and planet formation
We can test these hypotheses with the spectrally resolved line profiles that will be obtained by \OAS\ (Section~\ref{sec:distribution}).
An alternative interpretation of the low WISH fluxes is that the water in the disk atmosphere has a small filling factor, possibly because it is concentrated in rings similar to those found for the continuum emission from disks and in some molecular species observed with \citep{Long19,Andrews18b,Law21}. 
\OAS\ can investigate this scenario by linking observed water characteristics, e.g, temperature, emitting radius, with disk characteristics in other tracers across a large sample.

Finally, and most importantly, with \OAS\ we will be able to constrain the extent to which water has been concentrated into icy planet-building solids, answering questions raised by the WISH results. Ice abundances are difficult to measure directly, because most of the water ice resides in the cold disk midplane and is locked up in solids the size of pebbles, or perhaps even in larger planetesimals or protoplanets. These large solids do not produce a significant spectroscopic signature in the infrared. They are, moreover, too cold to emit in the phonon modes (at 45 \um\ and 62 \um), and because they do not overlap significantly along the line of sight to the near- to mid-infrared continuum (3-15 \um) for non-edge on viewing geometries, they will also not produce shorter wavelength absorption features (at 3 \um, 6.2 \um, and 11.5 \um). As a result, these solids are effectively ``hidden'' in the disk midplane. Furthermore, when icy solids grow beyond centimeter size, they are difficult to detect at any wavelength.

Although cold, planet-building water ice is difficult to measure directly, we can infer its mass indirectly by measuring the total disk gas mass 
(see Section 4.4).
and comparing it with the gaseous water vapor content of the disk. The total mass of water in the disk, in both icy and gaseous form, is expected to be approximately 0.2-0.6\% of the total gas mass. By measuring the difference between the expected total water mass and that of the measured water vapor, we can infer the mass in water ice. By further comparing the inferred total mass of water ice with the observed mass in small pebbles, as measured by the millimeter continuum emission of disks, we can infer the mass of water ice that has been converted into larger, planet-building solids that are not directly observable.

With its high sensitivity, \OAS\ may also be able to measure the location of the midplane water snowline in disks. 
Lines in the FIR/submm offer the best opportunity to probe this region (e.g., compared to the MIR wavelengths probed by JWST) because of their ability to penetrate the continuum optical depth and reach the midplane.
In the Solar System, the midplane water snowline divides the terrestrial and giant planets, with all of the large planets, the gas and ice giants, located beyond the snowline. Current planet formation models predict that most planets form beyond the midplane snowline (gas giants, ice giants, and possibly smaller planets) and may subsequently undergo radial migration and dynamical scattering \citep{Lin96,Rasio96}. Small water-rich bodies from larger distances (icy asteroids and comets) are the expected source of water for dry planets that form within the midplane snowline (e.g., Earth). Measurements of the location of the midplane snowline in disks, when compared with exoplanet demographics (orbital radii as a function of mass) will assess the role of the midplane snowline in determining the architecture of planetary systems, as well as illuminate the extent to which other processes alter exoplanetary orbital radii.

\OAS\ will obtain spectra for protoplanetary disks in nearby star-forming regions spanning a range in evolutionary stage (Class I, II, and flat spectrum sources), disk continuum flux, and stellar accretion rate. The measured line profiles of water and its isotopes, HD, and other gas tracers, will chart out the evolution in the water vapor distribution and gas content of disks, from which we will infer their progress toward planet building. The observations will also measure the location of the midplane water snowline in a subset of disks.

In its first few months of operation \OAS\ will carry out sensitive spectral scans across all four bands of approximately six disks spanning a range of stellar masses and evolutionary states, quantifying their spectral properties and making any needed modifications to the survey strategy; it will also observe an additional 50 or more disks in specific, well-chosen wavelength settings. Over the entire 1-year mission, \OAS\ will obtain similar observations of many more disks, including $\sim 20$ Class I sources with weak molecular envelopes and $\sim 100$ Class II and flat spectrum sources.

\subsection{Water abundance across evolutionary stages}\label{sec:waterabun}
\OAS\ will be the only facility capable of observing warm to cold water vapor, crucial to determining the total water content in disks. 
The abundance and distribution of water during planet formation remains poorly constrained, with few detections of water vapor in disks beyond the mid-infrared, which probes only the very surface region of the disk (its atmosphere) within the midplane snowline.
By measuring the water vapor abundance across evolutionary stages, \OAS\ will track the timescales over which water is locked into solid bodies. This in turn will determine how much water can be supplied to planets as they form, since terrestrial planets likely inherit their water from the solids while gas giants accrete much of their mass from the gas.

\jwst\ will significantly improve our understanding of water in the mid-IR. However, as noted by Decadal Question E-Q1c, additional {\it longer wavelength observations of cooler regions} of the disk are needed to understand disk composition and answer the question: \emph{How common is planetary migration, how does it affect the rest of the planetary system, and what are the observable signatures?} This relates to the method of using the ratio of volatile elements, e.g., carbon to oxygen, in an exoplanet's atmosphere to determine its formation location relative to major snowlines. Successful implementation of this techniques requires knowledge of how the radial distribution of water gas and ice evolves in protoplanetary disks. \OAS\ will observe the cold water vapor not probed by \jwst\ by targeting the ground state ortho and para transitions, collecting statistics on the cold water abundance in disks across evolutionary stages. 
In the largest disks, those spanning more than $\sim2$” on the sky, \OAS\ also has the potential to spatially resolve the emission from cold water vapor.

\begin{figure}[ht]
   \begin{center}
   \includegraphics[width=0.8\textwidth]{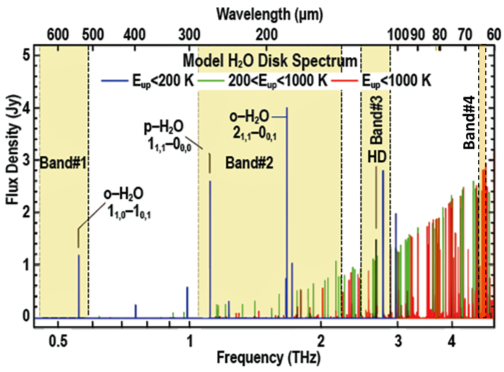}
   \end{center}
   \vspace{0mm}
  \caption{Model water spectrum for a protoplanetary disks. \OAS\ will measure the water vapor content of disks throughout the planet forming region using multiple water transitions that probe both the outer and inner disk, as well as intermediate regions near the snowline. Crucially, \OAS\ is able to observe the ground-state ortho and para water lines and the J=1-0 transition of HD simultaneously. \label{fig:spectra}}
\end{figure}

\OAS\ will target a suite of water lines spanning a large range in excitation energy, enabling measurements of the water content at larger disk radii, as well as lines with upper state energies 
$> 1000$~K 
that both originate near the snowline and are capable of penetrating the overlying disk continuum to  potentially reach
midplane, where most of the mass is located (Figure~\ref{fig:spectra}). Thus, \OAS\ will be able to observe water lines originating from many more regions in the disk than those accessible to \emph{JWST}.
With its large collecting area, \OAS\ will be far more sensitive than \Hers\ to the cold-to-warm water vapor in the disk atmosphere beyond 1 au, or roughly the midplane water snowline (Figure~\ref{fig:cartoon}). In a two hour integration \OAS\ will reach an rms sensitivity of 3 mK for $\mathrm{o-H_{2}O (1_{10}-1_{01})}$, an order of magnitude below the upper limits of the \Hers\ non-detections \citep{Du17}. Using the optically thin \watisoB\ emission lines, \OAS\ will measure water vapor abundances for disks ranging in age from less than 1 million years to greater than 10 million years. The large spectral range allows \OAS\ to measure water lines spanning a wide range of excitation energies, from 53 to 1729 K, and probe both warm (100-400 K) water vapor near the snowline and cold ($<100$ K) water vapor in the outer disk. Crucially, \OAS\ will observe the ground state ortho and para water lines, which cannot be observed with any other facility. 
Observations of both ortho and para transitions are necessary to determine the total water abundance. 
When combined with near-IR detections of hot water vapor in the inner disk atmosphere from \jwst, these measurements will provide a complete census of the water vapor in disks.

Along with simultaneous measurements of the total gas mass (Section~\ref{sec:hd}), these observations will also determine the fraction of the total water content that is in the form of water ice. These results will provide important constraints on the role of water ice in grain growth and on the solid water reservoir available to forming comets and planets.

\subsection{The spatial distribution of water vapor}\label{sec:distribution}
\begin{figure}
  \begin{minipage}[c]{0.5\textwidth}
    \includegraphics[width=\textwidth]{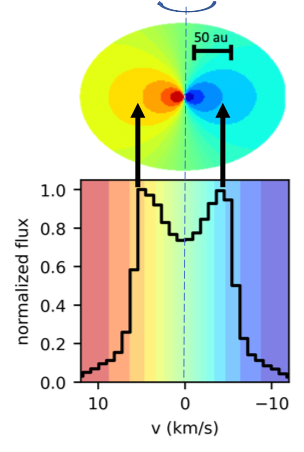}
  \end{minipage}\hfill
  \begin{minipage}[c]{0.5\textwidth}
    \caption{Flux normalized, disk integrated spectra from a protoplanetary disk model inclined 45 degrees from face-on  at 0.9 \kms\ spectral resolution. Emission from disks is Doppler shifted due to Keplerian rotation. Emission with a high velocity offset originates from small radii, while emission close to the systemic velocity originates from large radii. \OAS\ will leverage this to trace emission in spectrally resolved lines to the radius in the disk where the emitting molecules reside.} \label{fig:doppler}
  \end{minipage}
\end{figure}

\OAS\ will map the water vapor distribution toward a wide variety of disks, answering the question ``Where is the water?''
As stated in the NASA Astrophysics Roadmap, mapping the location of water in protoplanetary disks is crucial for understanding the transport of water during planet formation. Because disk rotation follows a Keplerian velocity profile, the radius at which gas emission originates changes the shape of the spectral line profile. The resulting Keplerian broadening of the spectral line profile dominates over other contributions to the line width, such as thermal and turbulent broadening. Thus, high spectral resolution observations of molecular lines in disks can be used to determine the radial location of the emission without having to spatially resolve the disk, a
technique known as doppler tomography or tomographic mapping. As shown in Figure~\ref{fig:doppler}, the velocity offset for emission originating in the inner disk is of order several \kms, assuming a disk inclination of 45 degrees, while in the outer disk the velocity offset is much smaller. The spectral resolution of \OAS\ is $<1$ \kms, which is easily able to distinguish emission originating in the inner versus outer disk.


Using tomographic mapping, \OAS\ will determine if water vapor is returning to the gas with the inward drift of icy dust grains, enriching the water content of the terrestrial planet forming region. Additionally, \OAS\ will probe the location of the midplane water snowline by observing multiple water lines with high ($\sim 1000$ K) upper state energies, which models predict to emit primarily from inside the midplane water snowline \citep{Rasio96,Notsu16}.
Thus, \OAS\ has the potential to make the first measurements in non-outbursting disks of the location of the
disk midplane water snowline,
an important disk landmark in the core accretion picture. Current planet formation models predict that most planets form beyond the snowline (gas/ice giants and possibly
smaller planets) and subsequently experience radial migration and dynamical scattering \citep{Lin96}.
These measurements, when compared with exoplanet
demographics (orbital radii as a function of
mass), will enable us to assess the role of water ice
in determining the architecture of planetary systems \citep[i.e.,][]{Kennedy08,Fernandes19} and illuminate the extent to which processes such as migration and dynamical scattering
alter exoplanetary orbital radii, addressing Decadal Question E-Q1c.

\begin{figure}[ht]
   \begin{center}
   \includegraphics[width=1.\textwidth]{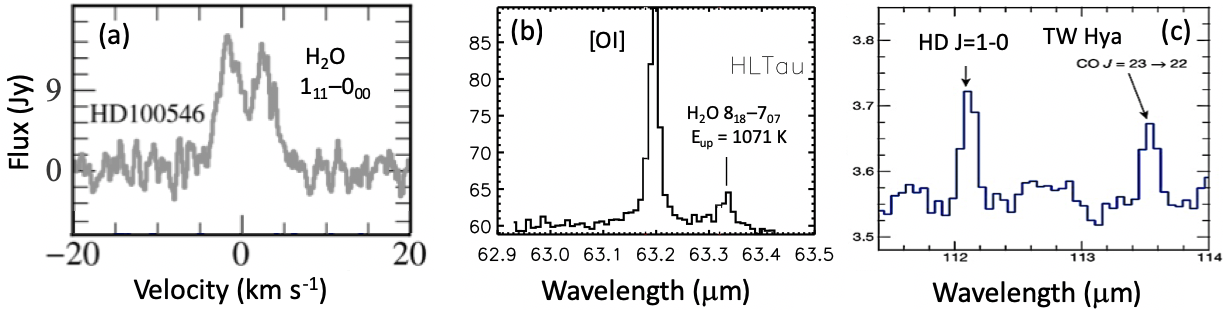}
   \end{center}
   \vspace{0mm}
  \caption{a) Herschel-HIFI detection of ground-state \wat\ emission \citep{vanDishoeck21}, tracing cold water vapor in the outer disk. b) Herschel-PACS spectrum \citep{Riviere12} showing unresolved lines targeted by \OAS\ Band-4. The high energy
\wat\ transition probes gas within the snowline. The strong atomic [OI] line could trace disk winds (Section~\ref{sec:distribution}). c) Herschel-PACS detection of the 
spectrally
unresolved HD J=1-0 line \citep{Berg2013}, a key measure of total disk mass (Section~\ref{sec:hd}). \OAS\ will provide high-sensitivity velocity-resolved observations of these and other transitions in more than 100 protoplanetary systems. \label{fig:FigD4}}
\end{figure}

During its baseline mission \OAS\ will explore
the role of disk winds in protoplanetary disks using
the 63 \um\ [OI] line (Figure~\ref{fig:FigD4}). Winds influence 
the evolution of the disk
through the removal of mass and angular momentum
from the disk. Observations of [OI] with Herschel
found that the majority of Class I sources, and
some Class II sources, had extended [OI] emission
suggesting large scale jets / outflows. With its high
spectral resolution, \OAS\ will disentangle the wind
component of [OI] from emission originating in the
disk and outflow.

\subsection{Measuring gas mass}\label{sec:hd}
\OAS\ will use HD to measure the gas mass in disks, allowing a comparison of relative \wat\ vapor abundance across systems. 
Protoplanetary disks — the sites for planet formation — will evolve in mass and size as more mass is accreted from the surrounding core, as angular momentum redistribution within the disk allows accretion onto the central star, and as disk winds return material to the molecular cloud. The disk gas mass is a fundamental property of the disk that regulates its dynamical evolution and its chemical evolution. The disk mass also controls the formation of planets through, e.g., coagulation of small dust grains into pebbles that settle to the midplane where planetesimal formation can ensue and gravitational instability of the gas disk can lead to a local concentration of mass. Observations reveal a general trend of gas rich disks at early evolutionary stages, with gas-to-dust ratios of $\sim100$, as seen in the ISM. Over time, the gas mass drops and the gas-to-dust ratio decreases until the dust mass exceeds the gas mass, known as the debris disk phase. However, the factors (e.g., disk characteristics and the molecular cloud environment) that control this evolution are not well understood.

Crucially for tracing the water, knowledge of the total disk mass is required to determine the total water (gas+ice) abundance, as chemical abundances are normalized to the hydrogen content. The column density of water vapor can be determined directly from observations of optically thin emission lines, however ice abundances derived from observations of broadband spectral features are rife with uncertainties \citep{McClure15}. An accurate determination of the disk mass is therefore needed to determine how much of the water vapor supplied to the disk has been converted to ice.

The main contributor to the disk mass is \hh, which does not emit for the majority of regions in the disk because the molecule has no permanent dipole moment and has large energy spacings that are not matched with the local temperatures. The ground-state transition is the quadrupole J = 2-0 transition with an energy spacing of 510 K. Thus, exciting an \hyd\ molecule to the J = 2 state requires high gas temperatures. \hyd\ emission therefore originates only from the illuminated surface layers of the disk within a fraction of an au of the central star. Most of the gas is at larger radii and is much colder. To determine the total gas mass then, alternative tracers must be used.

By far the most commonly used gas mass tracers in protoplanetary disks are continuum emission from dust and emission from rotational transitions of CO. Each method relies on different problematic assumptions. The uncertainties in using dust emission to determine gas mass fall into two categories: assumptions in converting from dust emission to dust mass, and in converting from dust mass to gas mass. Even when the full infrared spectral energy distribution is available, uncertainties in the dust grain optical properties and the grain size distribution lead to significant uncertainty in the derived dust mass from observed emission. Then, to convert from dust mass to gas mass, a gas-to-dust mass ratio must be assumed. Typically, this value is taken to be 100, as has been measured in the ISM. However, several factors can change this ratio in disks, including loss of gas due to disk winds and accretion onto the central star, which will decrease the gas-to-dust ratio, and growth of dust grains beyond cm sizes, at which point the dust emission is no longer observable. This will increase the apparent gas-to-dust ratio.

Additionally, assuming a constant gas-to-dust ratio across the disk is not appropriate. As dust grains grow in size they decouple from the gas, experiencing a headwind which results in a loss of angular momentum and inward radial drift. High spatial resolution observations at millimeter wavelengths demonstrate that the outer radius of the dust disk is often much smaller than the outer radius of the gas disk \citep{Huang18, Trapman19}.

The CO abundance relative to \hyd\ in the ISM is well constrained to be 5\ee{-5} to 2\ee{-4} \citep{Bergin17}.
However, when converting from CO abundance to \hyd\ in a protoplanetary disk additional corrections must be made to
account for the reduced abundance of CO relative to \hyd\ in the surface layer, where CO is photodissociated, and near the cold midplane, where CO is frozen out onto dust grains \citep{Miotello14}. Chemical reactions within the disk also destroy CO \citep{Schwarz18}. This reduction in the CO gas abundance varies from disk to disk and as a function of location within a single disk \citep{Zhang19}, introducing large uncertainties when converting to total gas mass.

\begin{figure}
  \begin{minipage}[c]{0.7\textwidth}
    \includegraphics[width=\textwidth]{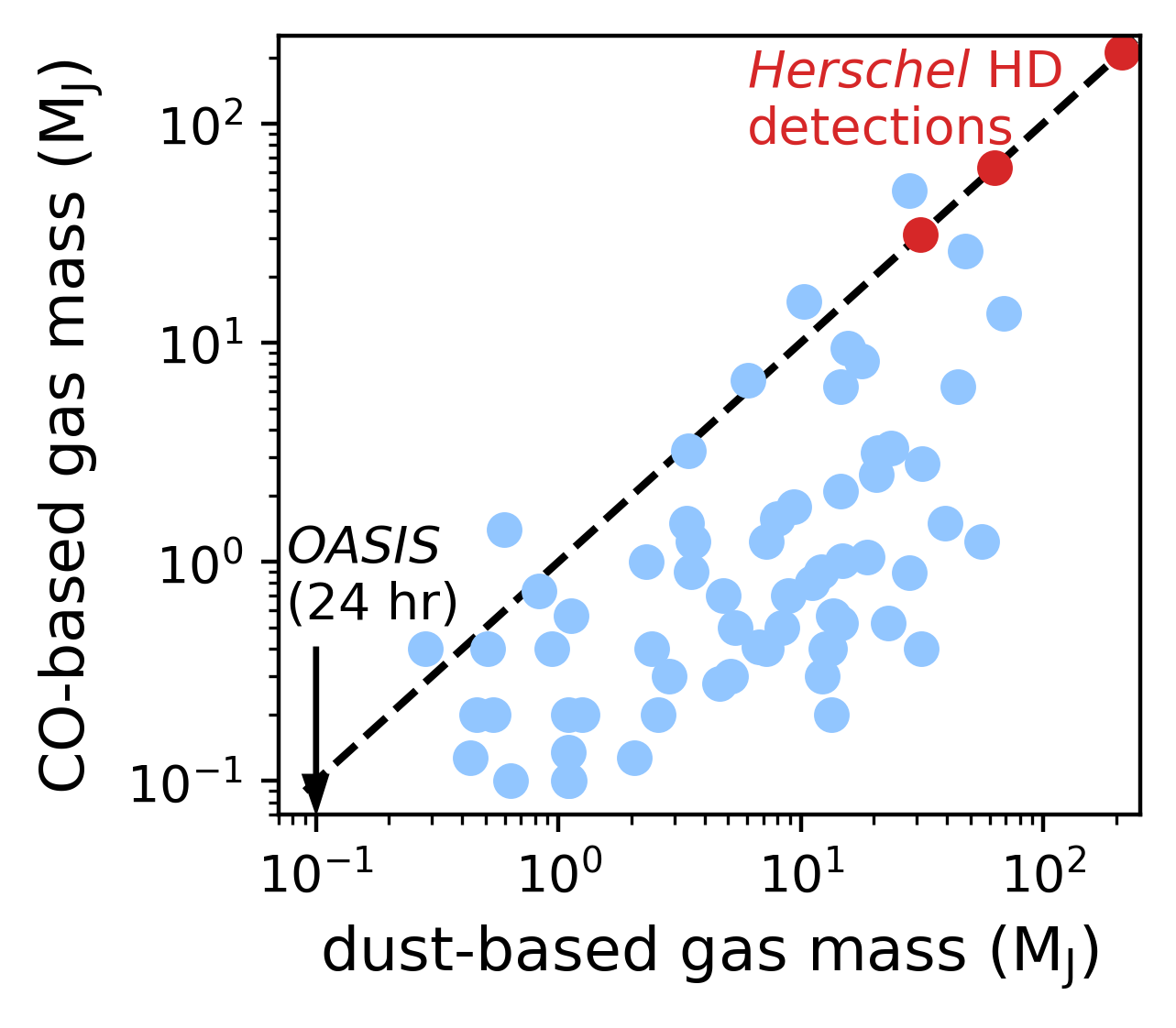}
  \end{minipage}\hfill
  \begin{minipage}[c]{0.3\textwidth}
    \caption{Blue dots show disparity between disk
mass estimates based on CO (y axis) and
dust continuum (x axis) observations with ALMA.
The two mass tracers rarely agree (dashed line),
with CO-based measurements, often giving a lower
mass. Red dots show the mass derived from the HD 1-0 line. HD observations allow a more direct measurement
of gas mass. Due to low integration times and limited sensitivity, \Hers\
only detected HD in three massive disks. \OAS\ will measure total gas mass in 100+
protoplanetary systems down to 0.1 Jupiter masses.} \label{fig:covsdust}
  \end{minipage}
\end{figure}

Given the myriad assumptions that go into each technique, it is not surprising that the two methods of determining gas mass rarely agree (Figure~\ref{fig:covsdust}). In fact, most observed disks appear to have significantly less gas than expected from the dust when using CO as the mass probe. Alternative mass probes, preferably requiring fewer assumptions, are needed to determine the true disk gas mass. One possibility is to use the disk rotation curve to constrain the enclosed mass. This technique has recently been demonstrated for a disk with a large total mass of $0.08\pm0.04$ solar masses \citep{Veronesi21}. However, because disks must always be less massive than the central star in order to remain gravitationally stable, the contribution of the disk to the rotation curve is small. Thus, this technique is only feasible for a small number of the most massive disks.


As an exciting alternative, the \hyd\ isotopologue HD can be used to trace disk mass while avoiding many of the limitations of other mass tracers. The HD abundance relative to \hyd\ is well constrained. HD emission is optically thin and not subject to the chemical processing that can change the CO abundance relative to \hyd\ \citep{Bergin17}. However, accurate determination of the gas mass still requires knowledge of the disk temperature structure, as HD does not emit appreciably below 20 K.

Near the end of its lifetime, Herschel targeted HD in seven systems, resulting in three detections \citep[Figure~\ref{fig:FigD4},][]{Bergin13,McClure16}. These were all massive disks, with the HD-derived gas masses of ~30-210 \mj. Crucially, these mass measurements reveal that both CO and \wat\ gas are depleted in these disks relative to the ISM \citep{Du15,Schwarz16}. No current observatory is capable of detecting HD. \OAS\ will measure the total gas mass in $\sim100$ protoplanetary systems, down to masses as low as 0.1 \mj. When combined with \OAS\ observations of cold water vapor, this determines the amount of water removed from the outer disk and placed into water ice in the planet forming midplane.

\subsubsection{A survey of protoplanetary disk mass}
\OAS\ will measure disk gas mass in a wide range of protoplanetary systems.
The high sensitivity of \OAS\ will allow it to detect HD at a 5-sigma level in disks down to 0.1 Jupiter masses with 24 hours on source, based on model fluxes \citep{Trapman17}. This is a factor of 300 more sensitive than what was achieved with \Hers. 
Converting the HD detections into an accurate total gas mass requires knowledge of the disk temperature structure. The design of \OAS\ allows for simultaneous observations in the four bands. Thus, while integrating on HD, \OAS\ will be able to observe multiple H$_2^{16}$O and $^{12}$CO lines spanning 55-1729 K in excitation energy, compared to 128.49 K for the HD J=1 excited state. These optically thick lines will provide direct measurements of the gas temperature throughout the disk. With \OAS’s high spectral resolution, doppler tomography can then be used to map temperatures to a physical radius in the disk. Thus, \OAS\ observations can be used to obtain measurements of the disk gas masses without the need for ancillary data.

At 2.675 THz, the frequency of the HD J = 1-0 line, the diffraction-limited beam size of \OAS\ is $\sim1.6$”. The largest protoplanetary disks have gas disks which subtend up to 10” on the sky. Thus, \OAS\ will be able to spatially map the  HD emission in protoplanetary disks for the first time. In unresolved disks with strong HD detections, doppler tomography of the HD spectral line can be used to similar effect. When paired with
temperature maps based on \watisoA\ and CO, this will greatly improve constraints on the temperature of the HD
emission and thus result in greater precision when determining the disk gas mass. Additionally, we will be able to map \hyd\ surface density profiles from the HD emission line profile — this is completely unique information.
Accurate surface density profiles are important for determining dynamics in protoplanetary disks, particularly
in the most massive systems, where local over-densities can drive instability, ultimately leading to the
formation of massive planets. The gas masses derived from HD will then be combined with the water vapor
abundances from \watisoB\ to determine the total (gas+ice) water content of protoplanetary disks.

\section{Baseline and Threshold Observations}
\label{sec:obs}
The limited number of cold water vapor detections
in protoplanetary disks to date are biased toward
the most massive systems \citep{Zhang13,Fedele13,Du17}. To
probe the timescale over which pebble growth / migration,
planetesimal formation, and giant planet formation
operate, we will study $>120$ disks in the one year Baseline \OAS\
mission and $\sim 40$ disks in the six month Threshold Mission,
spanning a range of evolutionary states, including
Class I (envelope + disk), Class II (disk only), and intermediate
flat-spectrum sources. Targets will cover
a range of expected disk masses and radial extents,
with the average disk 
being
far more compact than 
the large disks that are
typically studied. We will also study the impact of
stellar properties that are expected to influence water
abundance and distribution (spectral type, accretion
rate, and multiplicity). The large collecting area of
\OAS\ allows the first statistical study of water in protoplanetary disks spanning a large range of physical
properties. In the $\sim 10$ targets with gas disks 6 - 10" in diameter, \OAS\ will spatially resolve the distribution of cold water vapor for the first time.

Near the start of its mission, \OAS\ will carry out
deep spectral scans across all four bands toward
one to two Class 0/I sources, Class II sources, and flat spectrum sources
representative of the sample. The observed spectral
properties will be used to optimize the list of targeted
transitions in the broader sample. These observations
will be an order of magnitude more sensitive
than the deepest observations with Herschel. \OAS\
will then survey additional disks in specific, well-chosen
wavelength settings. Over its one-year baseline
mission, \OAS\ will obtain similar observations of
many more disks, including 20+ Class I sources with
weak molecular envelopes and 100+ Class II and
flat-spectrum sources. Based upon available models
and Herschel data, the expected integration time required
to achieve $\sim 5 \sigma$ detections toward the weakest
target line (HD) will be between $\sim1$ and 12 hours, depending
on disk mass and distance. \OAS’ unique
combination of sensitivity, spectral resolution, and 
broad spectral coverage is likely to produce many discoveries beyond the science described here.

\bmhead{Acknowledgments}
K.S. acknowledges support from the European Research Council 
under the European Union’s Horizon 2020 research and 
innovation program under grant agreement No. 832428-Origins.
J.B.B. acknowledges support from NASA through the NASA Hubble Fellowship grant No. HST-HF2-51429.001-A awarded by the Space Telescope Science Institute, which is operated by the Association of Universities for Research in Astronomy, Incorporated, under NASA contract NAS5-26555.

\bibliography{PPDscience.bib}

\end{document}